\documentclass{elsart}

\usepackage{graphicx}
\usepackage{amssymb}
\def \mdot {\dot{M}_{w}}

\begin{document}
\begin{frontmatter}



\title{Transient outburst mechanisms  \\ 
in Supergiant Fast X--ray Transients}


\author{Lara Sidoli}
\ead{sidoli@iasf-milano.inaf.it}

\address{INAF/IASF-Milano, Via Bassini 15, 20133 Milano (Italy)}

\begin{abstract} 
The recent discovery of a new class of recurrent and fast X--ray transient sources, 
the Supergiant Fast X--ray Transients, poses interesting questions on the possible 
mechanisms responsible for their transient X--ray emission. 
The association with blue supergiants, the spectral properties similar to those 
of accreting pulsars and the detection, in a few cases, of X-ray pulsations, 
confirm that these transients are High Mass X-ray Binaries. 
I review the different mechanisms proposed to explain their transient 
outbursts  
and the link to persistent wind accretors.
I discuss the different models in light of the new observational 
results coming from an on-going monitoring campaign of four Supergiant Fast X--ray Transients 
with \emph{Swift}.
\end{abstract}

\begin{keyword}
X--rays \sep individuals: IGR~J11215--5952, XTE~J1739--302, IGR~J17544--2619, IGR~J16479--4514, AX~J1841.0--0536/IGR~J18410--0535.
\end{keyword}

\end{frontmatter}

\section{Introduction}

An unusual X--ray transient in the direction of the
Galactic centre region, XTE~J1739--302, was discovered in 1997 
(Smith et al. 1998). 
It  displayed peculiar properties, particularly related with the duration
of the outburst: this source remained active only one day.
Since the X--ray spectrum was well fitted with a thermal bremsstrahlung 
with a temperature of $\sim$20~keV, resembling the
spectral properties of accreting pulsars, it was at first classified 
as a peculiar Be/X-ray transient with an unusually short outburst (Smith et al. 1998).

The INTEGRAL satellite has been performing a monitoring of the Galactic plane since 
its launch in October 2002 (Bird et al. 2007). 
Thanks to these observations, several other sources have been 
discovered, displaying similar properties to XTE~J1739--302: 
they show sporadic, recurrent, bright and short flares (with a typical duration of a few hours; 
Sguera et al. 2005, 2006;  Negueruela et al. 2006a). 
The X-ray spectra resemble the typical shape of High Mass X--ray Binaries (HMXBs) 
hosting X--ray pulsars,
with a flat hard power law below 10 keV, and a high energy cut-off at about 15-30~keV, sometimes
strongly absorbed at soft energies (Walter et al., 2006; Sidoli et al., 2006). 

Follow-up X--ray observations 
(e.g. Kennea et al. 2005, Tomsick et al. 2006) 
allowed to refine the source positions at the arcsec level, and to perform IR/optical
observations, which permitted to associate these transients with OB supergiant companions 
(e.g. Halpern et al. 2004, Pellizza et al. 2006, Masetti et al. 2006, 
Negueruela et al. 2006b, Nespoli et al. 2008).
These two main characterizing properties 
(the unusually short transient X--ray emission together with the association with blue 
supergiant companions) suggested that these sources define a new class of HMXBs, later named
Supergiant Fast X--ray Transients (SFXTs). 
Indeed, HMXBs with supergiant companions were previously 
known to show only persistent X--ray emission (e.g. Nagase, 1989).
Thanks to the observations of several new SFXTs performed with 
INTEGRAL (e.g., Sguera et al. 2005, 2006, 2007a; Negueruela et al. 2006a; 
Walter \& Zurita Heras 2007; Blay et al. 2008)
and, at softer energies, with 
\emph{Chandra} (in't Zand 2005), XMM-Newton (Gonzalez-Riestra et al. 2004; Walter et al. 2006) 
and archival observations with ASCA (Sakano et al. 2002), 
other important properties have been observed:
SFXTs display a high dynamic range, spanning 3 to 5 orders of magnitudes, 
from a quiescent emission
at 10$^{32}$~erg~s$^{-1}$ (characterized by a very soft spectrum, likely thermal; 
IGR~J17544--2619, in't Zand 2005; IGR J08408--4503, Leyder et al. 2007) 
up to a peak emission in outburst of 10$^{36}$--10$^{37}$~erg~s$^{-1}$.
At least two SFXTs are X--ray pulsars: IGR~J11215--5952 ($P_{\rm spin}=186.78\pm0.3$\,s,
Swank et al. 2007) and AX~J1841.0--0536 ($P_{\rm spin}$$\sim$4.7\,s, Bamba et al. 2001).
To date, eight are the confirmed members of the class of SFXTs 
(IGR~J08408--4503, IGR~J11215--5952, IGR~J16479--4514, XTE~J1739--302, IGR~J17544--2619, 
SAX~J1818.6--1703, IGR~J18410-0535/AX~J1841.0-0536, IGR~J18483--0311), 
with other $\sim$15 candidates (see e.g., the new INTEGRAL sources web page at 
http://isdc.unige.ch/$\sim$rodrigue/html/igrsources.html)
which
showed short transient flaring activity, 
but with no confirmed association with an OB supergiant companion.
The field is rapidly evolving, with an increasing number of  new transients 
discovered by instruments with a wide field of view (INTEGRAL/IBIS or Swift/BAT), 
so we expect that the whole class will grow in the near
future.

The first SFXT displaying periodic outbursts is IGR~J11215--5952 
(Sidoli et al. 2006, Smith et al. 2006) which undergoes an outburst about 
every  165~days (Romano et al. 2007b). 
This periodically recurrent X--ray activity is probably related to the orbital period of the system.
Thanks to the predictability of the outbursts, 
an observing campaign was planned in February 2007 with \emph{Swift} (Romano et al. 2007a) of
the fifth known outburst from IGR~J11215--5952.
Thanks to the combination of sensitivity and time coverage, \emph{Swift} observations 
are a unique data-set and allowed a
study of  this SFXT from outburst onset to almost quiescence. 
These observations demonstrated (see the \emph{Swift}/XRT lightcurve in Fig.~\ref{fig:igr11215}) that 
short duration and bright flares 
are actually part of a longer accretion phase at a lower level, lasting days, not only hours.

\section{Transient outburst mechanisms in SFXTs}

The main mechanisms proposed to explain the transient outbursts in 
SFXTs are based on the structure of the supergiant wind,
and can be divided into models based on (1)-spherically symmetric clumpy winds,
(2)-anisotropic winds, and those which deal with (3)-gated mechanisms
able to stop the accretion depending on the properties of both the companion wind and of 
the compact object (neutron star magnetic field and spin period).

\subsection{Outbursts from spherically symmetric clumpy winds}

One of the first hypotheses proposed to explain the transient X--ray emission from
these sources was the \emph{clumpy wind model}, 
where the short flares in SFXTs were supposed to be produced by the sporadic accretion 
of massive clumps which compose the blue supergiant companion wind (in't Zand 2005).
Indeed, in the last years more attention has been paid to a clumpy wind structure 
in early type stars (see, e.g., Lepine \& Moffat 2008; Oskinova et al. 2007 and references therein)

In the framework of the wind accretion (Bondi \& Hoyle 1944) which should be at work in this new class
of HMXBs, the X--ray luminosity depends on 
the wind mass loss rate, $\mdot$,  and on the relative velocity, $v_{rel}$, 
of the neutron star and the wind,
and is proportional to $\mdot$$v_{rel}$$^{-4}$ (Waters et al. 1989). 
This implies that large density and/or velocity wind contrasts produce 
large variability in the accretion X--ray luminosity, and thus can result 
in a very large dynamic range.
Walter \& Zurita Heras (2007) explain INTEGRAL observations of the bright flares from SFXTs
as due to the interaction of the compact object orbiting at $\sim$10 stellar radii
from the supergiant companion, accreting wind clumps with a mass
of 10$^{22}$--10$^{23}$~g, and with a density ratio between the clumps and the inter-clump medium
of 100--10,000.

Recently, Negueruela et al. (2008) proposed a revised version of this
hypothesis, based on the clumpy wind model by Oskinova et al. (2007).
These authors proposed two possible configurations for the SFXTs: 
(a)- a circular orbit just outside the region where the 
supergiant wind is denser (within about 2
stellar radii) or (b)- a wide eccentric orbit, which allows longer quiescent intervals.
In both the proposed geometries, 
since the neutron star orbits outside the region of the wind where the number
density of the clumps is higher, the probability to accrete a clump is very low.

In this framework, the difference between SFXTs and persistent supergiant HMXBs
mainly depends on the orbital separation, which is within 
2 stellar radii for persistent sources,
where the neutron star orbits inside the dense region of the wind, 
and much more than 2 stellar radii for SFXTs, because of 
a much lower probability to accrete a single clump.

\subsection{Outbursts from anisotropic winds: a preferential plane for the outflowing wind}

\begin{figure}[ht!]
\begin{center}
\includegraphics*[angle=270,scale=0.60]{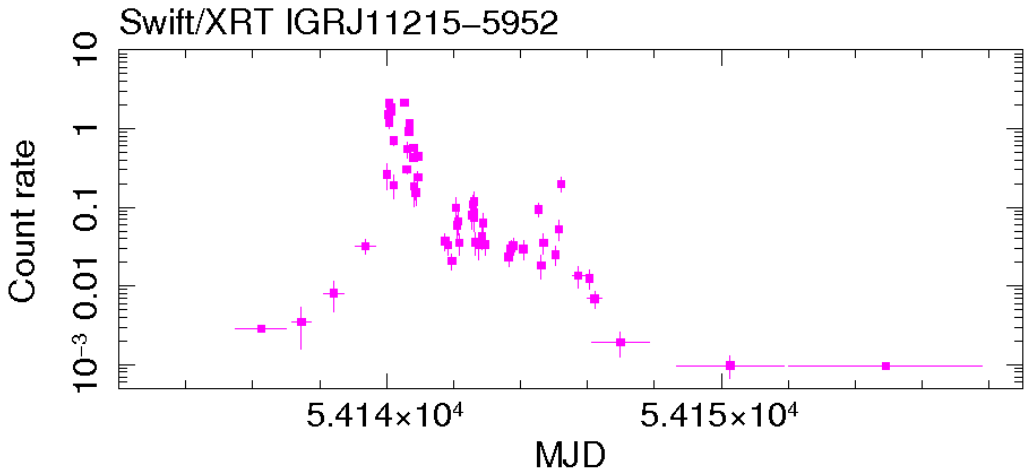} \\
\end{center}
\caption{\scriptsize Lightcurves of the SFXT IGR~J11215--5952 during the
outburst in February 2007 (MJD 54140=2007 February 9) observed with XRT in the energy range 0.2--10 keV.
}
\label{fig:igr11215}
\end{figure}

A second type of explanation involves \emph{anisotropic winds}, instead of spherically symmetric
winds (as assumed in the \emph{clumpy wind model}), and has been proposed to explain
the periodic outbursts observed from the SFXT IGR~J11215--5952 (Sidoli et al. 2007).
Thanks to the predictable outburst recurrence from this source, 
a monitoring campaign could be
planned with \emph{Swift}/XRT around the expected time of the new outburst, on 2007 February 9 
(Romano et al. 2007a), after 329 days from the previously observed outburst 
(Fig.~\ref{fig:igr11215}). The first periodicity discovered in the outburst
recurrence from this source was indeed 329 days (Sidoli et al. 2006; Smith et al. 2006).
For the first time a complete outburst
from a SFXT was observed from quasi-quiescence, up to the peak (lasting less than a day, with
several bright and short flares  and a high intensity variability), 
and then back to almost quiescence in a few days. 

The main important result of this campaign is that 
the outburst duration is much longer than a few hours, as was thought before
on the basis of INTEGRAL or RXTE observations. The hour duration flares seen by these 
intruments are indeed part of a much longer accretion phase, lasting a few days.
The periodic recurrence of the outburst suggests in a natural way the interpretation
of the recurrence timescale as the orbital period of the binary system, with the
outbursts triggered at, or near, the periastron passage.
Moreover, the shape of the X--ray lightcurve from the entire outburst can be hardly explained
with Bondi-Hoyle accretion (Bondi \& Hoyle, 1944) 
from a spherically symmetric wind in a binary system with an orbital
period of 329 days (Sidoli et al., 2007): the observed X--ray lightcurve 
is too narrow and steep compared with the
expected smoothly variable X--ray luminosity produced by accretion onto the surface
of an approaching neutron star along an eccentric orbit (even assuming very high
eccentricities).
We interpreted these observations as a clear evidence of the fact that the clumpy supergiant wind
is not spherically symmetric.
The short outburst can be explained with 
accretion from another wind component, for example an equatorially enhanced wind component 
(or any other preferential plane for the outflowing wind),
denser and slower than the symmetric polar wind from the blue supergiant, inclined with
respect to the orbital plane of the system, to explain the shortness of the outburst and the shape
of the source lightcurve. 
This geometry can also easily explain the periodic occurrence of the outbursts.

Based on the 329 days periodicity, we planned a second monitoring campaign with \emph{Swift} 
during the expected apastron passage (around 165 days after the last outburst). This new
campaign led to the observation of a new unexpected outburst from this source 
after a half of the previous period,  suggesting that $\sim$165 days 
is probably the new orbital period (Sidoli et al., 2007).
These new observations could be explained by two different geometries for the SFXT IGR~J11215--5952,
depending on the correct orbital period, P$_{\rm orb}$: 
(a)- if  P$_{\rm orb}$ $\sim$ 165 days, the orbit should be eccentric to produce only one 
outburst per orbit, when the neutron star crosses the equatorial wind; 
(b)- if P$_{\rm orb}$ $\sim$ 329 days, then the outbursts are two per orbit and the
orbit should be circular, since the two consecutive outbursts reach similar peak luminosities 
(Sidoli et al., 2007).

This interpretation could explain also the other SFXTs outbursts, although the geometry of this equatorially enhanced
wind component could be different from that explaining IGR~J11215--5952: for example, the equatorial wind 
could intersect the 
orbit of the compact object not near the periastron, but at different phases, thus producing
two regions on the orbit with enhanced wind density (where the outbursts are triggered), and thus two 
different periodicities, P$_1$ and P$_2$, in the outburst occurrence, if the orbit is eccentric, 
such that P$_1$+ P$_2$=P$_{\rm orb}$.
IGR~J11215--5952 could be probably the only system where the intersection of the equatorial outflow with the
neutron star orbit happens at (or very near to) the periastron. This could explain why it is more
difficult to discover a periodicity or (a double periodicity) in the outburst recurrence  in the other
SFXTs. 
Moreover, the thickness of this wind component could be different,
and the inclination with respect to the orbital plane could vary a lot among SFXTs, thus producing
very different phenomenologies, for example in the outburst  duration and recurrence, and covering
a different and variable fraction of the neutron star orbit.

There are indications that preferential planes in the outflowing supergiant winds
are probably present, especially from X--ray observations of other types of HMXBs.
In particular, apastron outbursts have been observed in supergiant HMXBs with known orbital periods
(e.g. Pravdo \& Ghosh 2001; Corbet et al. 2007)
where the presence of equatorially enhanced winds inclined with respect to the neutron star
orbit have been suggested in order to properly explain the observational X--ray data.

In conclusion, the observational evidence at the basis of our suggestion for the presence of 
a preferential plane for the outflowing wind in blue supergiant is provided by 
the periodicity in the IGR~J11215--5952 outburst recurrence and by the shape of the X--ray lightcurve, 
which cannot be 
produced by any temporary or transient feature in the stellar wind (such as wind streams or gas shells 
or shocks, or macro-structures of clumps), but must be due to some
permanent and orbital phase-fixed structure in the supergiant wind.  
This is naturally provided by a second wind component crossing the neutron star orbit.
We note that periodicities in the outburst recurrence begin to be found also in other SFXTs:
IGR~J18483--0311 displays a recurrence periodicity of $\sim$18.5 days (Sguera et al., 2007b);
this source was at first classified as a Be X--ray transient, but very recently Rahoui \& Chaty (2008)
optically identified the companion as a blue supergiant, thus the source is actually a SFXT;
IGR~J08408--4503 displays an X--ray activity (multiple flare duration and X--ray outburst recurrence)
which is compatible with an orbital period of 35 days (see Romano et al. 2009 for details).

\subsection{Gated mechanisms}

Other possibilities which have been suggested to explain the SFXTs outbursts, and
especially their high dynamic ranges, are based on gated mechanisms 
where the accretion is halted because of the presence of a magnetic
or a centrifugal barrier, which depends on the properties of the compact
object (the neutron star), in particular its spin period and its surface magnetic field.
Bozzo et al. (2008) applied this idea to SFXTs hosting a highly magnetized  neutron star 
[see also Grebenev \& Sunyaev  (2007) and references therein; Goetz et al. (2007)].
Wind accretion in a HMXB containing a magnetized neutron star depends on three different radii: 
accretion radius [R$_{\rm acc}$; Bondi \& Hoyle (1944)], corotation radius (R$_{\rm cor}$) 
and magnetospheric radius (R$_{\rm m}$).
While the corotation radius depends only on the neutron star mass and its spin period, 
the other two radii (R$_{\rm acc}$ and R$_{\rm m}$)
depend on the properties of the supergiant wind, while R$_{\rm m}$ also depend on 
the neutron star magnetic field, B 
(Illarionov \& Sunyaev 1975; Stella et al. 1986 and references therein).
Different types of interactions (accretion or inhibition of the accretion) 
between the  companion wind and the magnetized neutron star 
are possible depending on the relative positions of these three radii which, on the other hand,
are related to the neutron
star magnetic field and spin period, and to the properties of the supergiant wind 
(its velocity and density).
Direct accretion is possible only if R$_{\rm acc}$$>$R$_{\rm m}$ and R$_{\rm cor}$$>$R$_{\rm m}$.
In all the other cases, the accretion is prevented, by a magnetic barrier or by a centrifugal barrier.
Large luminosity swings ($\sim$10,000 or more) can result from transitions across these different regimes, 
which are produced if the relative positions of the three radii R$_{\rm acc}$, R$_{\rm cor}$, R$_{\rm m}$ change with
time, depending on a change in the wind parameters. 
Thus, modest variations in the wind
properties (as, for example, in a clumpy wind environment) 
can modify  R$_{\rm acc}$ and  R$_{\rm m}$ in such a way to produce an outburst.
This only applies if the three radii driving the accretion display very similar values.

Bozzo et al. (2008) show that an abrupt X--ray luminosity jump 
from the quiescence (at about 10$^{32}$~erg~s$^{-1}$) 
to the peak of an outburst in a SFXT (10$^{36}$--10$^{37}$~erg~s$^{-1}$) 
can be explained only if the the neutron star spins slowly
($\sim$1000~s or larger) and if it displays a magnetar-like magnetic field (B$\sim$10$^{14}$~G), in presence of
a relatively mild change in the wind density (one or two orders of magnitude, instead of the at least 4 orders
of magnitude requested by the clumpy wind model discussed above). 
Note that this model requires in any case a variability (although mild) in the wind parameters 
(such as clumps, or any other
structure in the wind with higher density than the smooth wind).
However density contrasts of 10$^{4}$
between the clumps and the inter-clump matter seem to be not unusual in supergiant winds 
[see, e.g., Oskinova et al. (2007)
and reference therein or the simulations of the line driven instability performed by Runacres \& Owocki (2005), 
which result in  clumps with density contrasts as large as 10$^{3}$--10$^{5}$ with respect to the inter-clump matter,
with a wind which remains clumpy far away from the hot star, out to 1,300 stellar radii].
Bozzo et al. (2008) applied this interpretation to the outburst observed in 2004 with
$Chandra$ from the SFXT IGR~J17544--2619 (in't Zand 2005) and show that the transition to the outburst
peak can be explained in terms of the magnetic barrier model, with a magnetar-like neutron star 
with a spin period of 1300~s (with a wind terminal velocity of 1400~km~s$^{-1}$, and an assumed orbital period 
of 10~days), if the wind mass loss rate changes from 10$^{-5}$~M$_\odot$~yr$^{-1}$ to  10$^{-4}$~M$_\odot$~yr$^{-1}$.

\section{Monitoring campaing with Swift}

In order to test the different proposed mechanisms for the transient outbursts, to measure the outburst
duration, to perform a truly simultaneous spectroscopy from soft to hard X--rays during outbursts,
 and to see in which status the SFXTs spend most of their lifetime, 
we have been performing a monitoring campaign of a sample of four SFXTs with \emph{Swift} since October 2007 
(Sidoli et al. 2008, Paper I):
the two prototypes XTE~J1739--302 and  IGR~J17544--2619, IGR~J16479--4514 
and the X--ray pulsar AX~J1841.0--0536/IGR~J18410--0535. 
The campaign consists of 2--3 observations/week/source (each observation lasts 1--2~ks).

\begin{figure}[ht!]
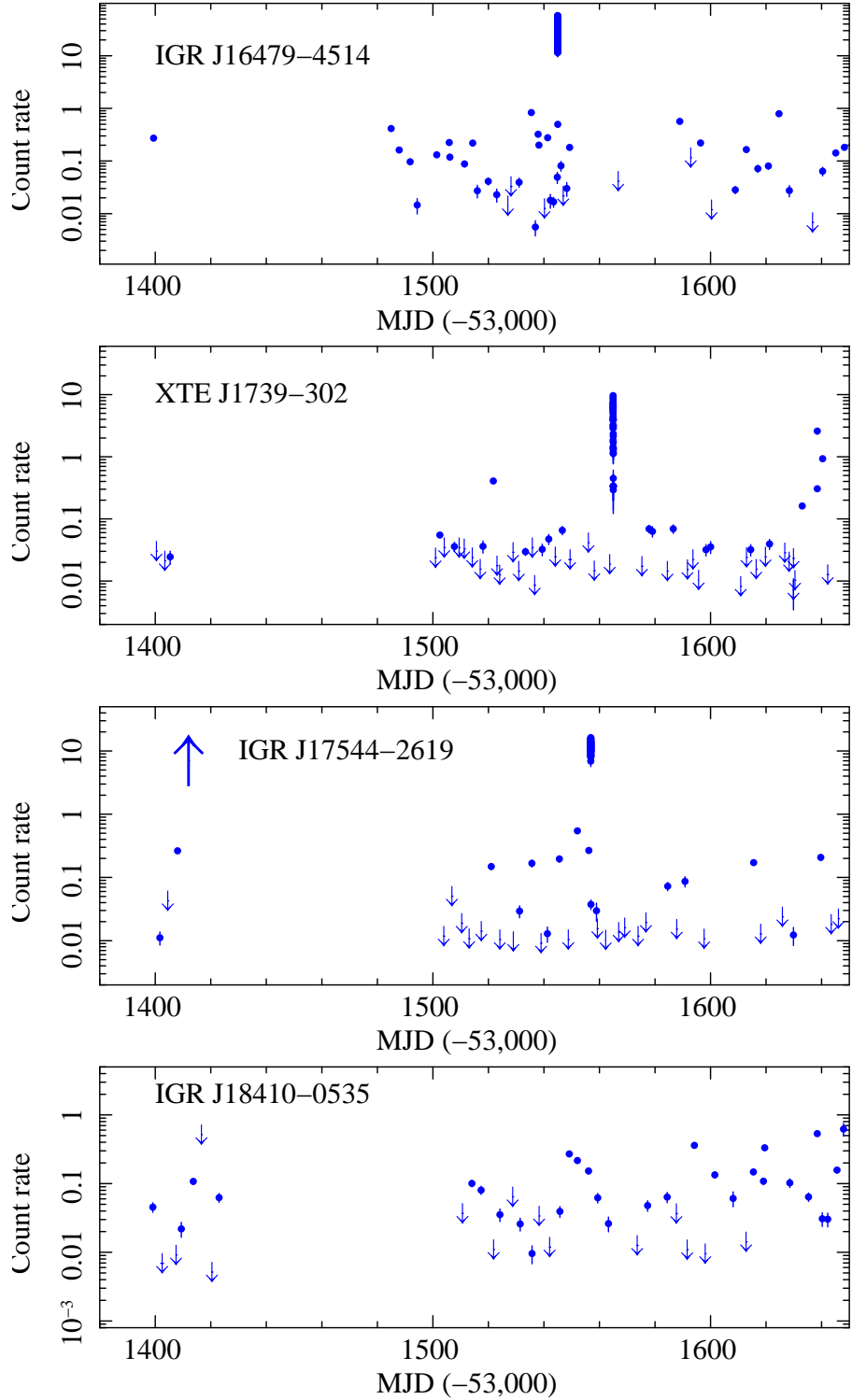

\begin{center}
\includegraphics*[angle=270,scale=0.5]{lsidoli_fig01.ps}
\includegraphics*[angle=270,scale=0.5]{lsidoli_fig02.ps}\\
\includegraphics*[angle=270,scale=0.5]{lsidoli_fig03.ps}
\includegraphics*[angle=270,scale=0.5]{lsidoli_fig04.ps}\\
\end{center}
\caption{\scriptsize Lightcurves of the 4 SFXTs monitored with \emph{Swift}/XRT (0.2--10 keV), from October 2007 to the
end of June 2008.The downward-pointing arrows are 3-$\sigma$ upper limits. 
The upward pointing arrow in the IGR~J17544--2619 lightcurve 
marks an outburst which triggered the BAT Monitor on MJD 54412 (Paper~I) but could not be observed with XRT because
of Sun-constraints.}
\label{fig:swiftlc}
\end{figure}

The results of this \emph{Swift} campaign are shown in Fig.~\ref{fig:swiftlc}, where the X--ray lightcurves
of the 4 SFXTs are displayed. The big gap in the observations between about December 2007 and February 2008
is due to the fact that the sources were Sun-costrained. 
The source fluxes are highly variable even outside the bright outbursts (which have been caught
in three of the four sources we are monitoring).
Variability on  timescales of days, 
weeks and months, is evident in the lightcurves, with a dynamic range (outside bright outbursts) 
of more than one order of magnitude in all four SFXTs.
The long term average behaviour of the source is a frequent low level flaring activity
with an average 2--10 keV luminosity of about 10$^{33}$--10$^{34}$~erg~s$^{-1}$ (Paper~I),
assuming the source distances reported in Rahoui et al. (2008).
The average spectra of the out-of-outburst emission is typically hard, well fitted with an absorbed power-law
with a photon index in the range 1--2.
The out-of-outburst emission in IGR~J16479--4514  and in AX~J1841.0--0536 appears to be modulated
with a timescale in the range 22--25~days, although we postpone a full timing analysis
to the end of the campaign.
These properties (flux variability of more than one order of magnitude {\em outside} the bright outbursts
together with the hard spectrum) demonstrate that SFXTs accrete matter even outside their bright outbursts,
and that the quiescence (characterized by a very soft spectrum and by a low level of emission
at about 10$^{32}$~erg~s$^{-1}$) is a much rarer state in these sources.

During the monitoring we caught bright outbursts from these sources, except from AX~J1841.0--0536,
which remarkably has not undergone an outburst for several months (maybe for years).
The first one was observed from IGR~J16479--4514 on 2008 March 19 
(Romano et al. 2008, Paper~II), which could be observed
over a wide energy band simultaneously with XRT and BAT, 
and exceeded at peak the flux of 10$^{-9}$~erg~cm$^{-2}$~s$^{-1}$. 
The 0.5--100 keV spectrum  can be adequately fit with models usually typical for accreting pulsars, an absorbed
powerlaw with a high energy cut-off, or a cutoff powerlaw, or 
a Comptonized emission (see Paper~II for details).
A time resolved spectroscopy of the flare resulted in similar fitting parameters for the X--ray emission,
with absorbing column density of $\sim$6$\times$$10^{22}$~cm$^{-2}$, 
a powerlaw photon index $\Gamma$$\sim$1,
a high energy cutoff E$_{\rm c}$$\sim$7~keV, and an e-folding energy E$_{\rm f}$$\sim$7--20~keV.
Assuming a distance of 4.9~kpc, the 0.5--100 keV luminosity  exceeds 5$\times$10$^{37}$~erg~s$^{-1}$.
The source lightcurve spanned at least 4 orders of magnitude in a few days monitored before
and after the bright flare.

A new outburst was observed also from IGR~J17544--2619 on 2008 March 31 (Sidoli et al. 2009, Paper~III).
Simultaneous observations with XRT and BAT allowed  to perform, in this case as well,
for the first time a broad band spectroscopy of the emission in outburst from this SFXT.
The deconvolution of the 0.3--50 keV emission results in a hard powerlaw-like spectrum below
10\,keV, with  high energy cut-off clearly emerging when fitting the BAT spectrum together with the XRT
data, typical of accreting pulsars (White et al. 1983).
A good fit is provided by a powerlaw with a high energy cut-off with the following bet-fit
parameters: $N_{\rm H}$=(1.1$\pm{0.2}$)$\times 10^{22}$ cm$^{-2}$, $\Gamma$=0.75$\pm{0.11}$,    $E_{\rm c}$=18$\pm{2}$~keV
and $E_{\rm f}$=4$\pm{2}$~keV, reaching a luminosity of  5$\times$10$^{37}$~erg~s$^{-1}$ (0.5--100~keV at 3.6~kpc).
The out-of-outburst emission observed with XRT below 10 keV 
appears to be softer and more absorbed than the emission during the flare.

The third outburst was caught from XTE~J1739--302 on 2008 April 8 (Paper~III). In this occasion,
two bright flares have been observed, separated by about 6000~s. 
The soft X--ray emission is much more absorbed than IGR~J17544--2619, showing a high absorbing
column density,  $N_{\rm H}$, of $\sim$1.3$\times$$10^{23}$~cm$^{-2}$, while the fit to the
broad band X--ray emission with a high energy cut-off
powerlaw resulted in the following parameters: $\Gamma$=1.4$^{+0.5} _{-1.0}$,    
$E_{\rm c}$=6$ ^{+7} _{-6}$~keV
and $E_{\rm f}$=16 $ ^{+12} _{-8}$~keV, reaching a luminosity of  3$\times$10$^{37}$~erg~s$^{-1}$ 
during the flare (0.5--100~keV, assuming a distance of 2.7~kpc).
The out-of-outburst emission observed with XRT below 10 keV 
is less absorbed than the emission during the flare, and displays similar spectral shape,
within the uncertainties. 

\begin{figure}[ht!]
\begin{center}
\includegraphics*[angle=0,scale=0.56]{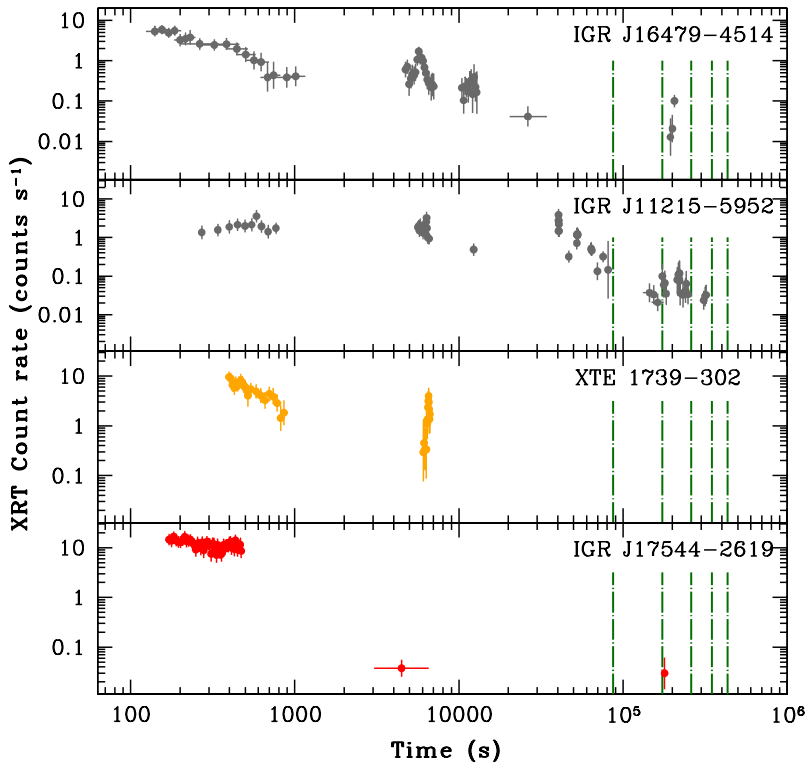} \\
\end{center}
\caption{\scriptsize Lightcurves of 4 SFXTs bright flares 
monitored with \emph{Swift}/XRT (0.2--10 keV): we show here the 2005 outburst from IGR~J16479--4514 (Paper~I),
the 2007 February outburst from the periodic SFXT IGR~J11215--5952 (Romano et al., 2007a), and the two 
flares from  IGR~J17544--2619 on 2008 March 31 and from XTE~J1739--302 on 2008 April 8 (Paper~III). Vertical dashed lines
mark different days, for clarity.}
\label{fig:swiftlc2}
\end{figure}

\section{Discussion and conclusions}

The monitoring campaing we are performing with \emph{Swift}  allow us to perform an interesting
comparison between the properties of the different SFXTs.

In Fig.~\ref{fig:swiftlc2} we show the lightcurves of 4 SFXTs. The duration and the shape
of the X--ray lightcurves appear to be similar, and so does the dynamical range spanned during the outbursts (about
three orders of magnitude). In the case of the April outburst from XTE~J1739--302 we could not follow entirely the
flare behaviour, but in the case of IGR~J16479--4514 and of IGR~J17544--2619 the outburst durations appear
very similar to that of the periodic SFXT IGR~J11215--5952, as the source dynamic ranges.
This demonstrates that it 
is no longer justified to exclude this source from the class of SFXTs \emph{only} because 
it displays
periodic outbursts.
This also demonstrates that the outburst duration in SFXTs is much longer than a few hours, 
as usually assumed
based only on INTEGRAL or RXTE observations, which of course could catch only the brightest part of the single flares.
Actually, these SFXT X--ray flares belong to a much longer accretion phase (outburst event) which last a few days.

The broad band X--ray emission displays a spectral shape very similar to that 
of accreting X--ray pulsars,
with a flat powerlaw hard X--ray emission below 10 keV with a high energy cutoff. 
Although no cyclotron lines have been observed yet in this new class of sources, the broad band 
spectral shape
(e.g. the cutoff energies) are consistent with a  pulsar magnetic field around 
a few 10$^{12}$~G, assuming
the empirical correlation (although never theoretically confirmed) 
between the high energy cutoff in X--ray pulsars and the observed cyclotron energies 
(Coburn et al. 2002).
This property seems to be in contrast with the hypothesis suggested by Bozzo et al. (2008)
that SFXTs are magnetars.
Moreover, Bozzo et al. (2008)  suggest that, to get the large dynamic ranges observed in SFXTs,
these transients should host neutron stars with long spin periods (of the order of 1000~s or larger),
which is not observed in SFXTs, to date: indeed, four SFXTs are X--ray pulsars
with much shorter spin periods: IGR~J11215--5952 (187~s, Swank et al. 2007), 
AX~J1841.0--0536 (4.7~s, Bamba et al. 2001), IGR~J18483--0311 (21~s, Sguera et al. 2007b), 
IGR~J16465--4507 (228~s, Lutovinov et al. 2005).
The large dynamic range between the quiescence and the maximum luminosity at the peak of the flares
can, in any case, be explained already if the supergiant winds are clumpy with
density contrasts as large as 10$^{5}$ (e.g. Runacres \& Owocki 2005), thus making
the hypothesis of magnetars in SFXTs not actually needed.  

The low-level flaring activity we observe in SFXTs during the out-of-outburst emission
can be explained both in the anisotropic wind hypothesis, where it is easily accounted for
by the accretion from the faster and less dense polar wind component from the supergiant star,
and in the spherically symmetric clumpy wind, for example assuming a 
distribution of clump sizes or masses (although the mass spectrum of the wind clumps is unknown). 
In any case, the flares from SFXTs cannot be explained
by the accretion of a single clump, since the duration of the outburst is much longer than
only a few hours, but extends for several days. 
A spherically symmetric wind (although clumpy) cannot explain
the periodicity of the outburst recurrence in IGR~J11215--5952.
Based on the anisotropic wind hypothesis, the outbursts should  occur at preferential orbital phases,
and display some sort of periodicity (or double periodicity or quasi-periodicity) 
in the outburst recurrence.
Although several observations have already been performed, not only with \emph{Swift}
but also with INTEGRAL, we emphasize the fact that, if all the other SFXTs are indeed
similar to IGR~J11215--5952 (which displays periodic outbursts every $\sim$165~days),
the likely periodicity (or double periodicity) in the outburst recurrence should be
of order of  months. This implies that probably they have not been discovered
yet because of a lack of observations.

\vspace{1cm}

\normalsize
\textbf{Acknowledgements}\\
\scriptsize
I am grateful to all my collaborators who made this research possible, 
in particular to Pat Romano,
who shared with me the  excitement for every new SFXT outburst observed during the 
{\it Swift} monitoring.
I would like to thank Monica Colpi, who organized a ``Neutron Star Day'' in Milano 
in 2006, during which lots of new ideas came out and all this started.
I would like to thank Neil Gehrels, Dave Burrows and all 
the {\it Swift} team for making these observations possible,
in particular the duty scientists and science planners.
This work was supported by contracts ASI/INAF I/023/05/0, I/008/07/0 and I/088/06/0.

\vspace{1cm}
\normalsize
\textbf{References}\\
\scriptsize
Bamba, A., Yokogawa, J., Ueno, M., et al., 2001, PASJ, 53, 1179-1183 \\
Bird, A.J., Malizia, A., Bazzano, A., et al., 2007, ApJS, 170, 175-186 \\
Blay, P., Martinez-Nunez, S., Negueruela, I., et al., 2008, A\&A 489, 669-676 \\
Bondi, H., \& Hoyle, F., 1944, MNRAS, 104, 273-282 \\
Bozzo, E., Falanga, M., Stella, L., 2008, ApJ, 683, 1031-1044 \\
Coburn, W., Heindl, W.A., Rothschild, R.E., et al., 2002, ApJ, 580, 394-412 \\
Corbet, R., Markwardt, C., Barbier, L., et al., 2007, PThPS, 169, 200-203 \\
Goetz, D., Falanga, M., Senziani, F., et al., 2007, ApJ, 655, L101-L104 \\
Gonzalez-Riestra, R., Oosterbroek, T., Kuulkers, E., Orr, A., Parmar, A. N., 2004, A\&A, 420, 589-594 \\
Grebenev,  S.A., Sunyaev, R.A., 2007, AstL, 33, 149-158 \\
Halpern, J.P., Gotthelf, E.V., Helfand, D.J., et al. 2004, The Astronomer's Telegram, 289 \\
in't Zand, J.J.M., 2005, A\&A, 441, L1-L4 \\
Illarionov, A.F., \& Sunyaev, R.A., 1975, A\&A, 39, 185-195 \\
Kennea, J.A., Pagani, C., Markwardt,  C., et al. 2005, The Astronomer's Telegram, 599 \\
Leyder, J.-C., Walter, R., Lazos, M., et al., 2007,  A\&A 465, L35 \\
Lepine, S., Moffat, A.F.J., 2008, AJ, 136, 548-553 \\ 
Lutovinov, A., Revnivtsev, M., Gilfanov, M., et al., 2005, A\&A, 444, 821-829 \\
Masetti, N., Morelli, L., Palazzi, E., et al. 2006,  A\&A, 459, 21-30 \\
Nagase, F., 1989, PASJ 41, 1-79 \\
Negueruela, I., Smith, D.M., Reig, P., et al. 2006a, in ESA Special Publication, ed. A.Wilson, Vol. 604, 165-170 \\
Negueruela, I., Smith, D.M., Harrison, T.E., et al., 2006b, ApJ, 638, 982-986 \\
Negueruela, I., Torrejon, J.M., Reig, P., et al., 2008, AIPC, 1010, 252-256 \\
Nespoli, E., Fabregatm J., Mennickent, R.E., 2008, A\&A, 486, 911-917 \\
Oskinova, L. M., Hamann, W.-R., Feldmeier, A., 2007, A\&A, 476, 1331-1340 \\
Pellizza, L.J., Chaty, S., Negueruela, I., 2006, A\&A, 455, 653-658 \\
Pravdo, S.H., Ghosh, P., 2001, ApJ, 554, 383-390 \\
Rahoui, F., Chaty, S., Lagage, P.-O., et al., 2008,  A\&A, 484, 801-813 \\
Rahoui, F., Chaty, S., 2008, A\&A, 492, 163-166 \\
Romano, P., Sidoli, L., Mangano, V. et al., 2007a, A\&A, 469, L5-L8  \\
Romano, P., Mangano, V.,  Mereghetti, S., et al., 2007b, The Astronomer's Telegram, 1151 \\
Romano, P., Sidoli, L., Mangano, V., 2008, ApJ, 680, L137-140   (Paper~II)     \\
Romano, P., Sidoli, L., Cusumano, G., 2009, MNRAS, 392, 45-51 \\
Runacres, M.C., \& Owocki, S. P. 2005, A\&A, 429, 323-333 \\
Sakano, M., Koyama, K., Murakami, H., et al., 2002, ApJS, 138, 19-34 \\
Sguera, V., Barlow, E.J., Bird, A.J., et al. 2005, A\&A, 444, 221-231 \\
Sguera, V., Bazzano, A., Bird, A. J., et al. 2006, ApJ, 646, 452 -463 \\
Sguera, V., Bassani, L., Landi, R., et al., 2007a, A\&A, 487, 619-623 \\
Sguera, V., Hill, A. B., Bird, A. J., et al., 2007b,  A\&A, 467, 249-257 \\
Sidoli, L., Paizis, A., \& Mereghetti, S., 2006, A\&A, 450, L9-L12 \\
Sidoli, L., Romano, P., Mereghetti, S., et al.,  2007, A\&A, 476, 1307-1315 \\
Sidoli, L., Romano, P., Mangano, V., et al., 2008, ApJ, 687, 1230-1235 (Paper~I) \\
Sidoli, L., Romano, P., Mangano, V., et al., 2009, ApJ, 690, 120-127 (Paper~III) \\
Smith, D.M., Main, D., Marshall, F., et al. 1998, ApJ, 501, L181-L184 \\
Smith, D.M., Bezayiff, N., \& Negueruela, I., 2006, The Astronomer's Telegram, 766 \\
Stella, L., White, N. E., Rosner, R., 1986, ApJ, 308, 669-679   \\
Swank, J.H., Smith, D.M., Markwardt, C.B., 2007, The Astronomer's Telegram, 999 \\
Tomsick, J.A., Chaty, S., Rodriguez, J., et al. 2006, ApJ, 647, 1309-1322 \\
Walter, R., Zurita Heras, J., Bassani, L., et al. 2006, A\&A, 453, 133-143 \\
Walter, R., \& Zurita Heras, J., 2007, A\&A,, 476, 335-340 \\
Waters, 1989, L.B.F.M., de Martino, D., et al., 1989, A\&A, 223, 207-218 \\
White, N.E., Swank, J.H., Holt, S.S., 1983, ApJ, 270, 711-734 \\


\end{document}